\documentclass{article}
\usepackage{graphicx}
\usepackage{hyperref}

\begin{document}

\title{Tools for assessing and optimizing the energy requirements of high performance scientific computing software}

\author{Kai Diethelm\footnote{Corresponding author: e-mail diethelm@gns-mbh.com}\\
GNS Gesellschaft f\"ur numerische Simulation mbH\\ Am Gau\ss berg 2, 38114 Braunschweig, Germany}

\maketitle

\begin{abstract}
  \emph{Score-P} is a measurement infrastructure originally designed
  for the analysis and optimization of the performance of HPC
  codes. Recent extensions of \emph{Score-P} and its associated tools
  now also allow the investigation of energy-related properties and
  support the user in the implementation of corresponding improvements. 
  Since it would be counterproductive to completely 
  ignore performance issues in this connection, the focus should not
  be laid exclusively on energy. We therefore aim to optimize 
  software with respect to an objective function that takes into
  account energy \emph{and} run time.
\end{abstract}

\section{The basic problem and static tuning approa\-ches for its solution}

To satisfy the demands from the scientific computing
community, the established high performance computing centers provide
a large amount of massively parallel computing hardware. One of the
main challenges that the HPC centers have to face today
is the cost of the energy required to operate this hardware which
already amounts to about
30\%\ of the total cost of ownership of a current HPC system, with a
rising tendency \cite{greenhpc}. Thus HPC centers will likely force their users to
optimize their software with respect to its energy requirements. In
this paper, we provide a brief survey of recent developments in this area.

\begin{figure}
\begin{center}\includegraphics[height=70mm]{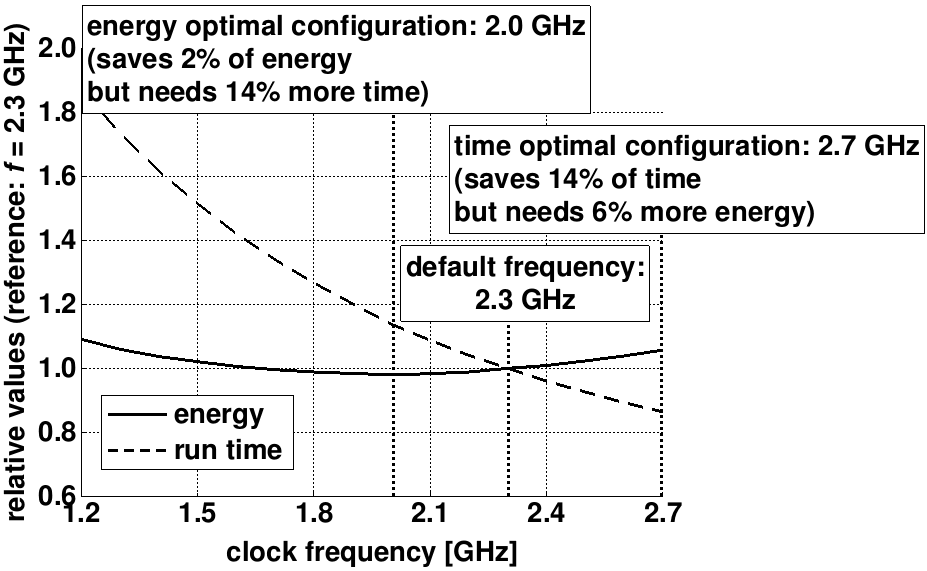}\end{center}
\caption{Relative energy requirements and run times for a typical run
  of the \emph{Indeed} finite element code on SuperMUC.}
\label{fig:indeedenergy}
\end{figure}

An early strategy implemented by certain HPC centers
\cite{lrz-energy} was to set the default CPU clock frequency of their
systems to a value much lower than the
highest possible frequency. This concept is based on the fact
that the total power required by a compute job
can be additively decomposed into a static component
$P_{\mathrm{st}} = \mathop{\mathrm{const}}$ (known as \emph{idle power}) and
a dynamic part that depends on $f$ and on the voltage $U$ as
$P_{\mathrm{dyn}} \sim U^2 \cdot f$ where, to obtain stability of the
operation, $U$ needs to be raised when $f$ is increased. 
The job's total energy is $E = \int_0^T
(P_{\mathrm st} + P_{\mathrm{dyn}}) dt$. An increase in
$f$  decreases the run time $T$, so the static
component of the total energy decreases while the dynamic component
may increase. Using a few test runs with typical input data sets, one
tries to gain an impression of the nature of this
dependence. For a specific example,
Fig.\ \ref{fig:indeedenergy} shows the result.

As expected, a moderate reduction of the clock frequency decreases
the energy consumption but also significantly
increases the run time. A straightforward application of the
idea thus implies that the available hardware cannot be fully utilized,
i.~e.\ the number of program runs that can be
executed during the system's life cycle is lower than it could
be. To justify the hardware investment,
it is thus not reasonable to focus
only on energy. A more useful metric is
the \emph{energy delay product} $\mathop{\mathrm{EDP}} =
E \cdot T^w$ where $E$ is energy, $T$ is run time, and $w$
is a parameter weighting energy and run time
according to the policy of the specific HPC center; typical
values are $w \in \{1, 2, 3\}$. An optimization with respect to such a
metric leads to a strategy that permits using higher
frequencies in spite of possibly larger energy requirements if the
run time savings are sufficiently large. Usually this is the case
for compute-bound software like, e.\ g., many finite element codes.
The CPU frequency is then fixed in advance for each code, and all
runs of this code are executed with this predefined
frequency. This approach is called \emph{static tuning}.

Apart from the CPU frequency, other parameters of a program run such
as, e.\ g., the number of OpenMP threads or the number of MPI processes,
can also be tuned in an analog way in order to optimize the energy
requirements.

\section{Energy analysis with \emph{Score-P} and the associated tools}

Tuning and optimization has been part of the scientific computing software
development process for a long time.
The focus of these activities has traditionally been on the
software's performance. A well established tool set for
this purpose consists of the automatic trace analyzer
\emph{Scalasca} (cf.\ \cite{scalasca2015} or \url{http://www.scalasca.org}), the interactive
trace analysis tool \emph{Vampir} (see \cite{BHJR:10:VampirOverview} or \url{http://www.vampir.eu}), the
profile analyzer \emph{CUBE} (cf.\ \cite{cubev4} 
or \url{http://www.scalasca.org/software/cube-4.x}), the profiling
and tracing system \emph{TAU} (see \cite{Shende:2006:Tau} or
\url{https://www.cs.uoregon.edu/research/tau}) and the
\emph{Periscope Tuning Framework} (PTF) (cf.\ \cite{AutoTune:Book2015} or \url{http://periscope.in.tum.de}) for
on-line analysis and tuning, and their underlying common measurement infrastructure
\emph{Score-P} (see \cite{knupfer2012score} or \url{http://www.score-p.org}). Recently
\cite{Score-E}, the systems have been extended so that they can now
also be used to analyze and optimize HPC codes with respect to
energy related metrics.

A key observation in such an analysis is that most codes exhibit 
dynamism, i.\ e.\ their behavior
varies over the run time. These variations can
be exploited for optimization purposes. The tools
listed above provide combined energy and performance
measurements for each slice of the run time. On this basis, one can 
develop and implement dynamic tuning strategies,
e.\ g.\ by adding commands to the code that change CPU frequencies,
the degree of parallelism, etc.\ as required in the current
situation. 

\begin{figure}
\begin{center}\includegraphics[height=70mm]{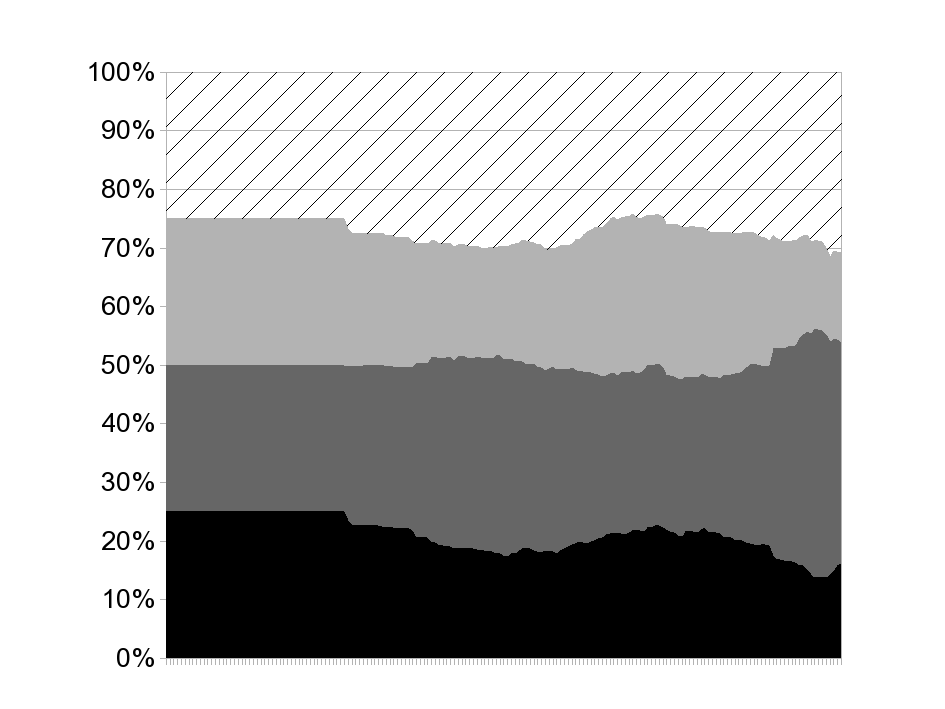}\end{center}
\caption{Distribution of work load over run time across four processes of a finite
  element simulation with domain decomposition and adaptive mesh.}
\label{fig:indeedloadbalance}
\end{figure}

A simple use case is a
domain decomposition based finite element simulation.
The simulation consists of a number of time steps;
each subdomain's mesh undergoes an adaptive refinement. Then it is
common for the workload to fluctuate between the processes
associated to the subdomains (see Fig.\ \ref{fig:indeedloadbalance}). 
By setting the clock frequency for each process according to its
current workload (i.\ e.\ according to the current number of elements
in its subdomain), processes with a smaller
workload can run at a slow clock speed, and hence require less energy,
but still finish their task in sync with the other
processes, so that the total run time is not negatively affected. 
Similar tuning actions can be implemented for, e.\ g., the
number of OpenMP processes, thus saving energy by temporarily switching
off some cores. Our tests indicate that static tuning
increases the energy efficiency (in the sense of
``energy-to-solution'') by some 10\%; we expect that dynamic tuning will lead to
improvements of about 30\%.

\section{Outlook: (Semi-)automatic dynamic tuning}

Our next steps in the development of the analysis and optimization
tools \cite{READEX-CSE-2015} aim at increasing the degree of automation. 
Specifically, a methodology has been derived that allows the user
to define points in the code at which a change of run time parameters
like CPU frequency of degree of parallelism is reasonable. Test runs
will then be used to find optimal values for these parameters in
certain situations defined, e.\ g., by the distribution of the
workload or other suitable data. During production runs of the code, a
runtime library will check the current situation at each
switching point, find the situation from the test runs that matches
best, and change the parameter set to the values identified as optimal
for this situation.
The implementation of this methodology is in progress. Our
goal is to achieve an improvent of the energy efficiency in the range
of 20\% to 25\% and to simultaneously reduce the programming effort 
in comparison to the manual dynamic tuning by 90\%.

\section*{Acknowledgement}
  The work described in this article was performed in the projects
  \emph{Score-E} \cite{Score-E} and \emph{READEX} (cf.\ \cite{READEX-CSE-2015} 
  or \url{http://www.readex.eu}).
  \emph{Score-E} is supported by the German Ministry of Education
  and Research under Grant No.\ 01IH13001A, and the project \emph{READEX}
  has received funding from the European Union's Horizon 2020 research and
  innovation programme under Grant Agreement No.\ 671657.

\end{document}